\documentclass[twocolumn,aps,prd,epsf]{revtex4}
\usepackage{epsfig}

\begin{document}

\title{Are the anti-charmed and bottomed pentaquarks molecular heptaquarks?}
\author{P. Bicudo}
\email{bicudo@ist.utl.pt}
\affiliation{Dep. F\'{\i}sica and CFIF, Instituto Superior T\'ecnico,
Av. Rovisco Pais, 1049-001 Lisboa, Portugal}
\begin{abstract}
I study the charmed $uudd\bar c$ resonance $D^{*-} p$ (3100) very recently discovered by 
the H1 collaboration at Hera. An anticharmed resonance was already predicted, in a recent 
publication mostly dedicated to the S=1 resonance $\Theta^+$(1540). To confirm these recent 
predictions, I apply the same standard quark model with a quark-antiquark annihilation 
constrained by chiral symmetry. I find that 
repulsion excludes the $D^{*-} p$ (3100) as a $uudd\bar c$ s-wave pentaquark. I
explore the $D^{*-} p$ (3100) as a heptaquark, equivalent to a $N-\pi-D^*$
linear molecule, with positive parity and total isospin
$I=0$. I find that the $N-D$ repulsion is cancelled by the
attraction existing in the $N-\pi$ and $\pi-D$ channels. In our framework this
state is harder to bind than the $\Theta^+$ described by a $k-\pi-N$
borromean bound-state, a lower binding energy is expected in agreement with the
H1 observation. 
Multiquark molecules $N-\pi-D$, $N-\pi-B^*$ and $N-\pi-B$ are also predicted.

\end{abstract}
\maketitle

%%%%%%%%%%%%%%%%%%%%%%%%%%%%%%%%%%%%%%%%%%%%%%%%%%%%%%%%%%%%%%%%%%%%%%%
%%                                                                   
%%
%%                                   SSS                             
%%
%%                                  S   S                            
%%
%%                                   S                               
%%
%%                                    S                              
%%
%%                                     S                             
%%
%%                                  S   S                            
%%
%%                                   SSS                             
%%
%%                                                                   
%%
%%%%%%%%%%%%%%%%%%%%%%%%%%%%%%%%%%%%%%%%%%%%%%%%%%%%%%%%%%%%%%%%%%%%%%%
\section{introduction}

\par
In this paper I study the anti-charmed $uudd\bar c$ resonance $D^{*-} p$ (3100) 
(narrow hadron resonance of 3099 MeV decaying into a $D^{*-} p$) very
recently discovered by the H1 collaboration at HERA 
\cite{H1}.
This extends to flavour SU(4) the SU(3) anti-decuplet 
\cite{Chemtob,Praszalowicz,Diakonov1}
which includes the recently discovered $\Theta^{+}$(1540)
\cite{Nakano,Barmin,Stepanyan,Barth,HERA-B}
and $\Xi^{--}$(1860)
\cite{Alt,Fischer,Price}. 
The $\Theta^+$(1540), $\Xi^{--}$(1860) and $D^{*-} p$(3100) are extremely 
exciting states, because they may be the first exotic hadrons to be discovered, 
with quantum numbers that cannot be interpreted as a quark and an anti-quark 
meson or as a three quark baryon. Exotic multiquarks are expected since the early
works of Jaffe
\cite{Jaffe,Strottman,Sorba,Roisnel},
and the SU(3) exotic anti-decuplet was first predicted within 
the chiral soliton model
\cite{Chemtob,Praszalowicz,Diakonov1}. Pentaquark structures
have also been studied in the lattice
\cite{Csikor,Sasaki,Chiu}.
However the nature of these particles, their isospin, parity
\cite{Hyodo}
and angular momentum, are yet to be determined.

\par
We recently completed a work on the $uudd\bar s$ $\Theta^+$ 
\cite{Bicudo00} 
where we indicate that the $\Theta^+$ is probably a 
$K-\pi-N$ molecule with binding energy of -30 MeV. 
In that work we first compute the masses of all the possible s-wave 
and p-wave $uudd \bar s$ pentaquarks, and we verify that these 
pentaquarks are hundreds of MeV too heavy to explain the 
$\Theta^{+}$ resonance, except for the $I=0$, $J^P=1/2^+$ state. 
However in this channel we find a purely repulsive exotic $N-K$ 
hard core s-wave interaction. This excludes, in our
approach, the $\Theta^+$ as a bare pentaquark $uudd\bar s$ state
or as a tightly bound s-wave $N - K$ narrow resonance.
We then add the $\pi - N$, $\pi - K$ and $N-K$
interactions to study the $\Theta^+$ as a borromean
three body s-wave boundstate of a $\pi$, a $N$ and a $K$
\cite{Bicudo00,Llanes-Estrada,Kishimoto}, with
positive parity 
\cite{Oh}
and total isospin $I=0$.
In that paper, and in a very recent work
\cite{Bicudo02},
we also address the S= -2, Q = -2 state $\Xi^{--}$, discovered
by the NA49 experiment with a mass of 1.862 GeV, 
indicating that this is a $\bar K - N- \bar K$ molecule with a binding 
energy of -60 MeV. 

In ref. 
\cite{Bicudo00} 
we also conclude suggesting the existence of similar anti-charmed 
$uudd\bar c$ and anti-bottomed exotic $uudd\bar b$ hadrons.
The anti-charmed pentaquark was widely expected
\cite{litera},
and its properties will certainly contribute to clarify the nature of the 
pentaquarks.
The state $D^{*-} p$ (3100) may be similar
to the $\Theta^+$, with the antiquark $\bar s$ replaced by a $\bar c$.
In this case it is natural to consider replacing the $K$ meson by a $D$ 
meson or by a $D^*$ meson, because the $D^*$ is also a narrow state.
For instance in the new positive parity $D_s$ mesons,
\cite{Bicudo01} 
the $\bar K- D$ and the $\bar K-D^*$ multiquarks are respectively candidates 
to the scalar  $D_s(2320)$ and to the axial $D_s(2460)$.
The energy of the is $D^{*-} p$ (3100) consistent with a  $D^*-\pi-N$ linear molecule 
with an energy of +15 MeV above threshold. This case differs from the previous ones
because here there is no negative binding energy. Nevertheless a system which 
energy is located slightly above threshold is still a narrow state, and in this
sense the $D^{*-} p$ (3100) remains in the same family of the $\Theta^+$
and of the  $\Xi^{--}$.

\par
Moreover a natural theoretical motivation exists for considering heptaquarks (or 
pentaquarks) and not just p-wave pentaquarks (or baryons) in the exotic multiplet 
of the $\Theta^+$. Supose that a given s-wave pentaquark hadron $H$
is studied and one concludes that it is unstable. Nevertheless 
one may consider that a flavor singlet quark-antiquark pair $u \bar u +d \bar d$ 
or $s \bar s$ is created in the hadron $H$. When the resulting heptaquark 
$H'$ remains bound, it is a state with an opposite 
parity to the original $H$, where the reversed parity occurs due to the intrinsic 
parity of fermions and anti-fermions. In this sense the new heptaquark $H'$ can be 
regarded as the chiral partner of $H$. And, because $H'$ is expected to be 
aproximately stable, it is naturally rearranged in a s-wave baryon in two 
s-wave mesons. The mass of the heptaquark $H'$ is expected to be slightly lower 
than the exact sum of these standard hadron masses due to the binding energy. 
This principle explains qualitatively the mass of the $\Theta^+$
\cite{Bicudo00}
for the $\Xi^{--}$ 
\cite{Bicudo02}
and the masses of the non-exotic multiquarks of the $D_s$ and $D^*_s$ familly
\cite{Bicudo01}.

\par
In this paper I extend the techniques used in our first publications 
to the $D^{*-} p$ (3100) and to the other similar narrow resonances
with an anti-charm or anti-bottom quark. 
A standard Quark Model (QM) Hamiltonian is assumed, with a
confining potential and a hyperfine term. Moreover the Hamiltonian
includes a quark-antiquark annihilation term which is the result
of spontaneous chiral symmetry breaking. I start in this paper by
reviewing the QM, and the Resonating Group Method (RGM)
\cite{Wheeler}
which is adequate to study states where several quarks overlap. 
Using the RGM, I show that the corresponding exotic baryon-meson short 
range s-wave interaction is repulsive in exotic channels and 
attractive in the channels with quark-antiquark annihilation. 
The short range repulsion contradicts the existence of narrow 
pentaquarks with an anti-charm or anti-bottom quark.  
I proceed with the study of the linear molecules or heptaquarks
$D-\pi-N$,  
$D^*-\pi-N$,  
$B-\pi-N$ and 
$B^*-\pi-N$. In particular the total energy of these systems is
discussed. Finally I conclude interpreting the $D^{*-} p$ (3100) 
and predicting related multiquarks.

%%%%%%%%%%%%%%%%%%%%%%%%%%%%%%%%%%%%%%%%%%%%%%%%%%%%%%%%%%%%%%%%%%%%%%%
%%                                                                   
%%
%%                                   SSS                             
%%
%%                                  S   S                            
%%
%%                                   S                               
%%
%%                                    S                              
%%
%%                                     S                             
%%
%%                                  S   S                            
%%
%%                                   SSS                             
%%
%%                                                                   
%%
%%%%%%%%%%%%%%%%%%%%%%%%%%%%%%%%%%%%%%%%%%%%%%%%%%%%%%%%%%%%%%%%%%%%%%%
\section{framework}

\par
Our Hamiltonian is the standard QM Hamiltonian,
\begin{equation}
H= \sum_i T_i + \sum_{i<j} V_{ij} +\sum_{i \bar j} A_{i \bar j} \,
\label{Hamiltonian}
\end{equation}
where each quark or antiquark has a kinetic energy $T_i$ with a
constituent quark mass, and the colour dependent two-body
interaction $V_{ij}$ includes the standard QM confining term and a
hyperfine term,
\begin{equation}
V_{ij}= \frac{-3}{16} \vec \lambda_i  \cdot  \vec \lambda_j
\left[V_{conf}(r) + V_{hyp} (r) { \vec S_i } \cdot { \vec S_j }
\right] \ . 
\label{potential}
\end{equation}
The QM of eq. (\ref{Hamiltonian}) 
reproduces the meson and baryon spectrum with quark and antiquark
bound-states (from the heavy quarkonium to the light pion mass).
The RGM was first applied by Ribeiro 
\cite{Ribeiro} 
to show that in exotic
$N-N$ scattering, the quark-quark potential together with
the Pauli repulsion of quarks explains the $N - N$ hard core
repulsion. Deus and Ribeiro
\cite{Deus} 
also showed that, in non-exotic channels, the 
quark-antiquark annihilation could produce a short core attraction. 
Recently, addressing a tetraquark system with the $\pi-\pi$ quantum 
numbers, it was shown that the QM also fully complies with the chiral 
symmetry, including the Adler zero  and the Weinberg theorem
\cite{Bicudo0,Bicudo1}.
Therefore the QM is adequate to address the anti-decuplet, which 
was predicted 
\cite{Chemtob,Praszalowicz,Diakonov1}
in an effective chiral model. 

\par
For the purpose of this paper the details of the potentials in eq. 
(\ref{Hamiltonian}) are unimportant, only its matrix elements matter. 
The hadron spectrum constrains the hyperfine potential,
\begin{equation}
\langle V_{hyp} \rangle \simeq \frac{4}{3} 
\left( M_\Delta-M_N \right)
\simeq M_{K^*}- M_K  \ .
\label{hyperfine}
\end{equation}
When a light quark is replaced by a heavy quark, say a
charmed quark, the hyperfine interaction is decreased, and 
it must also be replaced by 
$\langle{V_{hyp}}_D \rangle \simeq M_{D^*}- D_K$ .
The quark-antiquark annihilation potential $A_{i \bar j}$ is 
also constrained when the quark model produces spontaneous chiral 
symmetry breaking
\cite{Bicudo3,Bicudo4}.
The annihilation potential $A$ is present in the $\pi$
Salpeter equation,
\begin{equation}
\left[
\begin{array}{cc}
2 T + V & A \\
A & 2T +V
\end{array}
\right]
\left(
\begin{array}{c}
\phi^+ \\
\phi^-
\end{array}
\right) =
M_\pi
\left(
\begin{array}{c}
\phi^+ \\
-\phi^-
\end{array}
\right)
\label{pion BS}
\end{equation}
where the $\pi$ is the only hadron with a large negative energy
wave-function, $\phi^- \simeq \phi^+$.  In eq. (\ref{pion BS}) the
annihilation potential $A$ cancels most of the kinetic energy and
confining potential $2T+V$. This is the reason why the pion has a very
small mass. From the hadron spectrum and using eq. (\ref{pion BS})
the matrix elements of the annihilation potential are determined,
\begin{eqnarray}
\langle 2T+V \rangle_{S=0} &\simeq& {2 \over 3} (2M_N-M_\Delta)
\nonumber \\
\Rightarrow \langle A \rangle_{S=0} &\simeq&- {2 \over 3} 
(2M_N-M_\Delta)
\ ,
\label{sum rules}
\end{eqnarray}
where this result is correct for the annihilation of $u$ or $d$ quarks.

\par
The RGM 
\cite{Wheeler}
computes the effective multiquark energy 
using the matrix elements of the microscopic quark-quark interactions. 
Any multiquark state can be decomposed in combinations of simpler 
colour singlets, the baryons and mesons. 
%
%
%%%%%%%%%%%%%%%%%%%%%%%%%%%%%%%%%%%%%%%
%%                                   %%
%%                FFFFFF             %%
%%                F                  %%
%%                FFF                %%
%%                F                  %%
%%                F                  %%
%%                                   %%
%%%%%%%%%%%%%%%%%%%%%%%%%%%%%%%%%%%%%%%
\begin{figure}[t]
\epsfig{file=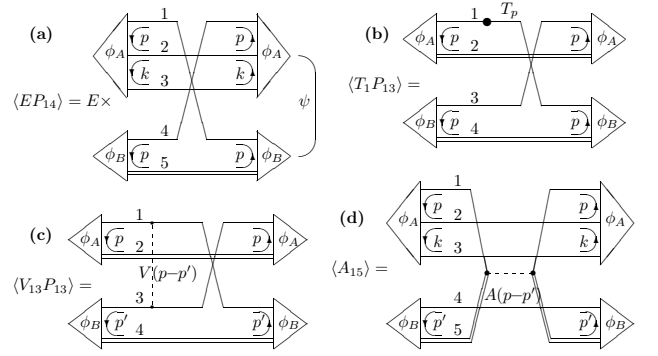,width=8.5cm} \caption{Examples of
RGM overlaps are depicted, in (a) the norm overlap for the meson-baryon
interaction, in (b) a kinetic overlap the meson-meson interaction,
in (c) an interaction overlap the meson-meson interaction, in (d)
the annihilation overlap for the meson-baryon interaction.}
\label{RGM overlaps}
\end{figure}
The wave functions of quarks are
arranged in anti-symmetrized overlaps of simple colour singlet
hadrons. 
Once the internal energies $E_A$ and $E_B$ of the two hadronic
clusters are accounted, 
\begin{equation}
{\langle \phi_b \phi_a | H \sum_p (-1)^p P |\phi_a \phi_b \rangle
\over
\langle \phi_b \phi_a | \sum_p (-1)^p P |\phi_a \phi_b \rangle
} = E_a+E_b +V_{a \, b} \ ,
\end{equation}
where $\sum_p (-1)^p P$ is the anti-symmetrizer, the remaining energy of the meson-baryon
or meson-meson system is computed with the overlap of the
inter-cluster microscopic potentials,
\begin{eqnarray}
V_{\text{bar } A \atop \text{mes } B}
&=& \langle \phi_B \, \phi_A |
-( V_{14}+V_{15}+2V_{24}+2V_{25} )3 P_{14}
\nonumber \\
&& +3A_{15} | \phi_A \phi_B \rangle /
\langle \phi_B \, \phi_A | 1- 3 P_{14} | \phi_A \phi_B \rangle
\nonumber \\
V_{\text{mes } A \atop \text{mes } B}
&=& \langle \phi_B \,
\phi_A | (1+P_{AB})[ -( V_{13}+V_{23}+V_{14}+V_{24})  
\nonumber \\
&& \times P_{13} +A_{23}+A_{14} ]| \phi_A \phi_B \rangle 
\nonumber \\
&& / \langle \phi_B \, \phi_A | 
(1+P_{AB})(1-  P_{13}) | \phi_A \phi_B \rangle
\ ,
\label{overlap kernel} 
\end{eqnarray}
where $P_{ij}$ stands for the exchange of particle $i$ with
particle $j$, see Fig. \ref{RGM overlaps}. 
It is clear that quark exchange provides the necessary colour octets to match 
the Gell-Mann matrices $\lambda_i$ present in the potential $V_{ij}$.
This results in eq.
(\ref{hyperfine}) or eq. (\ref{sum rules}) times an algebraic
colour $\times$ spin $\times$ flavour factor and a geometric
momentum overlap
\cite{Ribeiro3}.

\par
A good approximation for the wave-functions of the ground-state
hadrons is the harmonic oscillator wave-function,
\begin{equation}
\phi_{000}^\alpha(p_\rho) =  {\cal N_\alpha}^{-1}
\exp\left({ - {{p_\rho}^2 \over 2 \alpha ^2}}\right) \ , \ \
{\cal N_\alpha} = \left({ \alpha \over 2 \sqrt{\pi}}\right)^{3 \over 2} 
\ ,
\label{basis}
\end{equation}
where the inverse hadronic radius $\alpha$ can not be estimated 
by electron-hadron scattering because it is masked by the vector
meson dominance. $\alpha$ is the only free parameter in this framework.
In the case of vanishing external momenta $p_A$ and $p_B$,
the momentum integral in eq. (\ref{overlap kernel}) is simply
${\cal N_\alpha}^{-2}$.

\par
The annihilation potential only occurs in
non-exotic channels. Then it is clear from eq. (\ref{sum rules}) 
that the annihilation potential provides an attractive (negative) 
overlap. 
The quark-quark(antiquark) potential is 
dominated by the hyperfine interaction of eq. (\ref{hyperfine}), and 
in s-wave systems with low spin this  results in a repulsive interaction.
These results are
independent of the details of the quark model that one chooses to
consider, provided it is chiral invariant.
Therefore I arrive at the attraction/repulsion criterion,
\\
- {\em whenever the two interacting hadrons have quarks (or antiquarks)
with a common flavour, the repulsion is increased by the Pauli principle,
\\
- when the two interacting hadrons have a quark and an
antiquark with the same flavour, the attraction 
is enhanced by the quark-antiquark annihilation}.
\\
In the particular case of one nucleon interacting with anti-kaons 
and with kaons, this implies that the short range exotic $D-N$, $D^*-N$,  
$B-N$ and $B^*-N$ interactions are repulsive. This shows that the I=0
s-wave pentaquarks $uudd\bar c$ and $uudd\bar b$ are certainly quite
unstable. Higher isospin or spin systems are certainly more unstable in our 
framework because they have a higher repulsion. 
On the other hand the short range 
$\pi-N$,  $\pi-D$,  $\pi-D^*$,  $\pi-B$ and  $\pi-B^*$ interactions 
can be attractive. This motivates the study of a linear molecule
with a $N$, a $\pi$ and a $D$, or a $D^+$, or a $B$, or a $B^*$. 
Quantitatively
\cite{Bicudo00,Bicudo1,Bicudo2,Bicudo5},
the effective potentials computed for the different channels, are
\begin{eqnarray}
V_{D-N}&=& {1 \over 2}
{{1\over 2}+{1 \over 3}\vec \tau_D \cdot \vec \tau_N \over
\frac{3}{4}-\frac{1}{3}  \vec \tau_D \cdot \vec \tau_N}
\langle V_{hyp} \rangle \, {\cal N_\alpha}^{-2} 
\nonumber \\
&& + {1 \over 2}
{{1\over 2}+{1 \over 3}\vec \tau_D \cdot \vec \tau_N \over
\frac{3}{4}-\frac{1}{3}  \vec \tau_D \cdot \vec \tau_N}
\langle {V_{hyp}}_D \rangle \, {\cal N_\alpha}^{-2} \ ,
\nonumber \\
V_{D-N \atop \rightarrow D^*-N}&=& 
{{1+2\sqrt{3} \over 8}+{1 +\sqrt{3} \over 3}\vec \tau_D \cdot \vec \tau_N \over
\sqrt{3}+{1\over4 }+{5 \over 3  }  \vec \tau_D \cdot \vec \tau_N}
\langle V_{hyp} \rangle \, {\cal N_\alpha}^{-2} 
\nonumber \\
&& + {-{1\over 8 }-{4 \over 3 }\vec \tau_D \cdot \vec \tau_N \over
\sqrt{3}+{1\over4 }+{5 \over 3  }  \vec \tau_D \cdot \vec \tau_N}
\langle {V_{hyp}}_D \rangle \, {\cal N_\alpha}^{-2} \ ,
\nonumber \\
V_{D^*-N}&=& {1 \over 2}{
2+{7 \over 3}\vec \tau_D \cdot \vec \tau_N 
\over
{11 \over 4}+{7 \over 3}  \vec \tau_D \cdot \vec \tau_N}
\langle V_{hyp} \rangle 
\, {\cal N_\alpha}^{-2} 
\nonumber \\
&& + {1 \over 2}{ {-1\over 2}+{5 \over 3}\vec \tau_D \cdot \vec \tau_N 
\over
{11 \over 4}+{7 \over 3}  \vec \tau_D \cdot \vec \tau_N}
\langle {V_{hyp}}_D \rangle 
 \, {\cal N_\alpha}^{-2} \ ,
\nonumber \\
V_{\pi-N}&=& -{1 \over 3} \,  \vec \tau_\pi \cdot
\vec \tau_N \, \langle A \rangle  \, {\cal N_\alpha}^{-2} \ ,
\nonumber \\
V_{\pi-D}&=& -{4 \over 9} \, \vec \tau_\pi \cdot
\vec \tau_D \, \langle A \rangle  \, {\cal N_\alpha}^{-2} \ ,
\nonumber \\
V_{\pi-D^*}&=& -{4 \over 9} \, \vec \tau_\pi \cdot
\vec \tau_D \, \langle A \rangle  \, {\cal N_\alpha}^{-2} \ ,
\label{zero p}
\end{eqnarray}
where ${\vec \tau}$ are the isospin matrices, normalized with ${\vec \tau \,}^2= \tau(\tau+1)$. 
The vanishing momentum case of eq. (\ref{zero p}) is sufficient to compute 
the scattering lengths with the Born approximation. 
However the study of binding needs the finite momentum case. In the exotic 
$D - N$ and $D^* - N$ channels, it can be proved that the geometric result 
${\cal N_\alpha}^{-2}$ is then replaced by the separable interaction 
$ | \phi_{000}^\alpha \rangle \langle  \phi_{000}^\alpha |$.
In the non-exotic $\bar \pi - N$, $\pi - D$  and $\pi - D^*$ channels the
present state of the art of the RGM does not allow a precise
determination of the finite momentum overlap. Nevertheless I 
assume for simplicity the same separable interaction. This is
a reasonable approximation, because the overlaps
decrease when the relative momentum of hadrons $A$ and $B$
increases. Moreover when the hadronic potential $V_{AB}$ is in a separable
form $v|\phi_1 \rangle \langle \phi_1|$, the energy of the bound state and 
the matrix element of the potential are simple to compute. A boundstate 
coincides with a pole in the T matrix at a negative energy,
\begin{eqnarray}
&& T=  |\phi_1> {v \over 1 - {g_0}_{11} \, v }  <\phi_1| \ ,
\nonumber \\ 
&& {g_0}_{ij}=<\phi_i| 
{ 1 \over E + i \epsilon - {p^2 \over 2 \mu} }|\phi_j> \ ,
\label{first separable}
\end{eqnarray}
therefore one just has to find the energy that cancels $1 - g_0 \, v $.
This method also allows the computation of the matrix element of the potential
in the boundstate,
\begin{equation}
<V_{AB}> = v { {g_0}^2 \over \sum_j  {{g_0}_{j1}}^2 } \ .
\label{second separable}
\end{equation}
When $|\phi_1> $ is the harmonic oscillator state
of eq. (\ref{basis}), the necessary condition for binding is, 
\begin{equation}
-4 \mu \, v \ge \alpha^2 .
\label{third separable}
\end{equation}

%%%%%%%%%%%%%%%%%%%%%%%%%%%%%%%%%%%%%%%%%%%%%%%%%%%%%%%%%%%%%%%%%%%%%%%
%%                                                                   
%%
%%                                   SSS                             
%%
%%                                  S   S                            
%%
%%                                   S                               
%%
%%                                    S                              
%%
%%                                     S                             
%%
%%                                  S   S                            
%%
%%                                   SSS                             
%%
%%                                                                   
%%
%%%%%%%%%%%%%%%%%%%%%%%%%%%%%%%%%%%%%%%%%%%%%%%%%%%%%%%%%%%%%%%%%%%%%%%
\section{binding flavour $uudd \bar Q$ multiquarks }

\par 
The simplest pentaquarks are not 
expected to bind due to the attraction/repulsion criterion. For instance
the $D^{*-} p$ (3100) cannot be the ground-state $uudd\bar c$ pentaquark 
because the elementary color singlets $(uud)-(d\bar c)$ or  $(udd)-(u\bar c)$ 
are repelled, since the elementary color singlets share the same flavour $u$ 
or $d$. 
This also implies that the $D^{-} p$ and $D^{0} n$  systems are unbound.
The $uudd\bar c$ pentaquarks with spin, flavour or angular momentum
excitations will have larger masses and also large widths.
Nevertheless the $\bar c$ pentaquarks are more subtle than the
$\bar s$ ones, because the $D^*$ is quite stable when compared with 
the $K^*$. Therefore one should also consider to excite the spin
in the $l \bar c$ cluster, and this amounts to study $D^*-N$
bound-states. Indeed the effective potential of eq. (\ref{zero p})
is attractive in this case. However this state is coupled to the
$D-N$ case also in eq. (\ref{zero p}). Once the coupled channel
hamiltonian is diagonalized, the energy of the
$D^*-N$ is lifted and the attraction is essentially lost.
Therefore the simplest way to have attraction, together with a low energy and 
with a narrow width, consists in adding at least one quark-antiquark pair to 
the system. 

\par
Then this amounts to include a pion in the system, which can be attracted both
by the $N$ baryon and by the $D$ or $D^*$ meson to produce a $N-\pi-D$ 
or  $N-\pi-D^*$ linear molecule. 
For instance the flavour includes combinations of terms like 
$uud-d \bar u- u \bar c$ 
where the anti-quark $\bar u$ in the pion can be annihilated both by the 
$u$ present in the $N$ and by the $u$ present in the $D$. According to the
attraction/repulsion criterion this produces an attractive interaction.
The quark $d$ present in the nucleon cancels only part of the attraction to
the pion. Incidently the pion-nucleon $I=1/2$ attraction is fixed by
chiral symmetry, see reference
\cite{Bicudo2}.

\par
The proposed system $N-\pi-D$ and  $N-\pi-D^*$ are similar to the model 
for $\Theta^+$(1540) advocated in reference
\cite{Bicudo00},
in the present case the anti-quark $\bar s$ is replaced by a heavy
$\bar c$ or $\bar b$. The increase of the quark mass does not affect 
directly the attraction, where the $\bar Q$ is just a spectator.
However the size of the wave-functions $1/\alpha$ is affected. For instance
in an harmonic oscillator potential $\alpha$ is proportional to 
$^4\hspace{-.2cm}\sqrt{\mu}$,
and the reduced mass $\mu$ doubles when one changes from a light-light
meson to a heavy-light meson. This amounts to an increase of nearly 20\%
of the  $\alpha$ in the $D$ or $B$ meson. Because the $\alpha$ parameter
is increased only in one of the Jacobi coordinates, the average
$\alpha$ in  $\pi-D$ or $\pi-D^*$ or $\pi-B$ or $\pi-B^*$ is only 
expected to suffer a 10\%. This increase of $\alpha$ will decrease
effectively the attractive interaction. for instance the condition for 
a 2-body binding in eq. (\ref{third separable}) shows that this is
equivalent to decrease the strength of the attractive potential
by a factor of 20\%. Similar results are obtained in different
models of confinement, say in the funnel interaction which is
more adequate for heavy quarks.
In what concerns the repulsive $D-N$ potential, it is decreased
in the same way. Moreover the strength of the hyperfine potential is further 
decreased because $\langle {V_{hyp}}_D \rangle << \langle V_{hyp} \rangle $. 
For example the strength of the repulsive $D-N$ potential is expected
to decrease by 30 \%.

\par
I now use an adiabatic Hartree method to study the stability of the
linear $N -\pi-D$ molecule and related molecules with a
$D^*$, a $B$ or a $B^*$. Essentially the wave-function
of the pion is centered between the nucleon and the $D$, where the
nucleon and the $D$ don't overlap with each other. This results in
a linear molecule. for simplicity I use an averaged mass for the nucleon
and $D$ and for the pion interaction with these quark clusters. 
I solve a Schr\"odinger equation for the nucleon in the potential produced 
by a pion placed at the origin and by the other heavy-light meson placed at a 
distance $-{ \bf a}$ of the pion. 
The potential of the pion is produced by the $D$ meson at the point
$-{ \bf a}$ and the nucleon ${+\bf a}$. 
This produces three binding energies $E_D \, E_{\pi}, \, E_N$,
and three wave-functions. In the Hartree method the total energy is the 
sum of these energies minus the matrix elements of the potential energies,
This is easily computed once the two Schr\"odinger equations are solved,
with eqs. (\ref{first separable}), (\ref{second separable}) and 
(\ref{third separable}). The total energy is a function of the distance 
$\bf a$, and I minimize it as a function of a. The same steps are repeated
for the $N -\pi-D^*$, $N -\pi-B$ and $N -\pi-B^*$ systems.
At the point I am not yet able to bind these linear molecule systems,
with a negative binding energy. The same happened for the $N-\pi-K$ 
when we studied the $\Theta^+$.
\cite{Bicudo00}
Nevertheless the picture of a $K-\pi-N$ with a binding energy of $30$ MeV 
is still plausible because the medium range interaction remains to be used.
Therefore binding or near-binding is also plausible in the  
$N -\pi-D$, $N -\pi-D^*$, $N -\pi-B$ and $N -\pi-B^*$ systems.

%%%%%%%%%%%%%%%%%%%%%%%%%%%%%%%%%%%%%%%%%%%%%%%%%%%%%%%%%%%%%%%%%%%%%%%
%%                                                                   
%%
%%                                   SSS                             
%%
%%                                  S   S                            
%%
%%                                   S                               
%%
%%                                    S                              
%%
%%                                     S                             
%%
%%                                  S   S                            
%%
%%                                   SSS                             
%%
%%                                                                   
%%
%%%%%%%%%%%%%%%%%%%%%%%%%%%%%%%%%%%%%%%%%%%%%%%%%%%%%%%%%%%%%%%%%%%%%%%
\section{Conclusion and outlook}

I find that $N -\pi-D$, $N -\pi-D^*$, $N -\pi-B$ and $N -\pi-B^*$ nearly
bound I=0 linear s-wave molecules, and positive
parity are plausible. The absorption of a low energy pion and the 
resulting decay into a low energy p-wave $N-D(D^*,B,B^*)$ results in a 
narrow decay width. The $N -\pi-D^*$ and $N -\pi-B^*$ can also 
decay respectively into the three body systems $N -\pi-D$ and  $N -\pi-D$,
but this is again narrow since the $N -D^*$ and $N -B^*$ overlaps are
suppressed.
Moreover $N -D^*$, and $N -B^*$ pure I=0 s-wave, negative parity, pentaquarks 
are not completely ruled out, but they couple strongly with the decay channels
$N -D$ and $N -B$ (repulsive systems), and this results in a larger width.

\par
Comparing with the $K-\pi-N$ and assuming that it is a bound system,
I expect the $N -\pi-D^*$ and  $N -\pi-B^*$ to be
the most bound systems of the family studied in this paper, with a 
similar binding energy. 
Nevertheless the $N -\pi-D^*$ is less bound than the $N-\pi-K$,
due to a weaker $\pi-D(D^*,B,B^*)$ attraction when compared with the
$\pi_K$ attraction. Thus an energy of 14 MeV above threshold, 
corresponding to a $uudd\bar c$ mass of 3.099 GeV as observed by the
the H1 collaboration is plausible. The corresponding mass of the
multiquark with flavour $uudd\bar b$ is of the order of 6.416 GeV.
I what concerns the goundstate $N -\pi-D$ and $N -\pi-B$ the binding
energy is predicted to be higher because in this case the nucleon-
heavy-light meson repulsion is larger. So these states would have an
energy some MeV larger than respectively 2.957 GeV and 6.370 Gev.

\par
A more quantitative computation of the masses, sizes, and decay rates
of the proposed heptaquarks, including coupled channels and exact
three body computations will be done elsewhere. I expect that the
most relevant contributions that remain to be included in this framework
are the solution of the full three body relativistic Fadeev
equations, and the inclusion of the medium range interaction.
The medium range interaction, which in nuclear physics is described
by the sigma meson exchange, is equivalent to the coupling to
channels with multiple pions.

\acknowledgments
I thank Gon\c{c}alo Marques for discussions on the algebraic computations 
of this paper. I am also grateful to Katerina Lipka
and to Achim Geiser for discussions on the status of 
the experimental evidence of the $D^*-p$(3100).

%bbbbbbbbbbbbbbbbbbbbbbbbbbbbbbbbbbbbbbbbbbbbbbbbbbbbbbbbbbbb
%bbbbbbbbbbbbbbbbbbbbbbbbbbbbbbbbbbbbbbbbbbbbbbbbbbbbbbbbbbbb
%bb
%bb
%bbbbbbbbbbbbbbbbbbbbbbbbbbbbbbbbbbbbbbbbbbbbbbbbbbbbbbbbbbbb
%bbbbbbbbbbbbbbbbbbbbbbbbbbbbbbbbbbbbbbbbbbbbbbbbbbbbbbbbbbbb

\end{document}